\documentstyle[11pt,fleqn]{article}
\def\ref{par\noindent\hangindent=6mm\hangafter=1}
\begin{document}
\vbox{
%\rightline{}
\rightline{gr-qc/9406017}
\rightline{Nuovo Cimento A 107, 1543-1547 (1994)}
}
\baselineskip 8mm
%paper #5/1994

\begin{center}
{$S^1\times S^2$ as a bag membrane and its Einstein-Weyl
geometry}

\bigskip

H.C. Rosu\footnote{Electronic mail:
rosu@ifug.ugto.mx}
%\cite{byline}

{\it Instituto de F\'{\i}sica de la Universidad de Guanajuato, Apdo Postal
E-143, Le\'on, Gto, M\'exico}

\end{center}

\bigskip
\bigskip

\begin{abstract}

In the hybrid skyrmion in which an Anti-de Sitter bag is imbedded into
the skyrmion configuration a $S^1\times S^2$ membrane is lying on
the compactified spatial infinity of the bag [H. Rosu, Nuovo Cimento
B 108, 313 (1993)].
 The connection between the quark degrees
of freedom and the mesonic ones is made through the membrane, in a way
that should still be clarified from the standpoint of general relativity and
topology.
The $S^1 \times S^2$ membrane as a 3-dimensional manifold is at the same
time a Weyl-Einstein space. We make here an excursion through the mathematical
body of knowledge in the differential geometry and topology of these spaces
which is expected to be useful for hadronic membranes.

\end{abstract}

\vskip 1cm

%%%%%%%%%%%%%%%%%%%%%%%%        THE PAPER        %%%%%%%%%%%%%%%%%%%%%%%%
%%%%%%%%%%%%%%%%%%%%%%%%%  written by H.C. Rosu  %%%%%%%%%%%%%%%%%%%%%%%%%%%
%%%%%%%%%%%%%%%%%%%%%%%%%%%   1992 - Trieste   %%%%%%%%%%%%%%%%%%%%%%%%%%%% 

%%%%%%%%%%%%%%%%%%%%%%%%%%%%%%%%%%%%%%%%%%%%%%%%%%%%%%%%%%%%%%%%%%%%%%%%%
\section{Introduction}

We considered recently a hybrid skyrmion model possessing an
 Anti de Sitter minibag \cite{ro92}. In such a case, quarks could be
 interpreted as Dirac singletons, and are confined by
 geometry alone. However, when the spatial infinity of the
  Anti-de Sitter space is taken into account more carefully,
  one will discover, via a bosonization procedure, the possibility
  of communication across the boundary of the bag. This is reminiscent
  of the so-called Cheshire Cat principle which was much
  popularized by the Nordita group \cite{nnz85}.
   In fact, taking quarks and
  gluons to be indeed Dirac singletons, one could think of them as
  degrees of freedom belonging entirely to the membrane \cite{ff}.
   The scope of the present paper
  is to put together some geometrical aspects of the spatial
  infinity of Anti de Sitter spacetime, i.e., the $S^{1}\times S^{2}$
  membrane. This manifold is one of the eight classes of 3-dimensional
  geometries in Thurston classification \cite{t}. The most important
  properties are its Weyl-Einstein character and the fact it has no
  Einstein metric. One might say that $S^{1}\times S^{2}$ is a natural
  linking between the Einstein metric of the bag internal region
  and the Weyl geometry reigning all over the outside.

Much of the basic mathematical facts collected here can be found in
the excellent book of Besse \cite{be87}. Also the 1989 Durham review
of Tod \cite{t90} is a very useful reference. Further mathematical
progress may be found in the Princeton studies of Guillemin \cite{g}.

\section{Einstein spaces}
\subsection{Inside the AdS bag}
The spacetime within the bag is the AdS vacuum solution of Einstein
equations. It is a vacuum solution of constant curvature characterized
by one real number only, the scalar curvature. For physicists it is one
of the most celebrated Einstein spaces.

Generally an Einstein metric is defined to satisfy ,
\begin{equation}
Ric(g)=cg
\end{equation}
where c is some constant (if dim $ M\ge 3$)
The differentiability of the metric in Einstein spaces may be
changed by isometries, and to obtain optimal smoothness, harmonic
coordinates are required.
 The first theorem to be mentioned is that of
 DeTurck-Kazdan \cite{dk81}: {\it
 one could find an atlas
in any Einstein manifold in dim $ \ge 3$ with real analytic transition
functions, so that the metric is analytic in each coordinate chart}.

One of the best known theorems concerning Einstein manifolds is, as a
matter of fact, a corollary of DeTurck-Kazdan theorem and has to do with
local isometric embeddings. It states that: {\it any Einstein manifold
 (M,g)
of dimension greater or equal to three is locally isometrically
embeddable in ${I\!\!R} ^{n(n+1)/2}$}. The embeddable dimension
 is high but at
the present time it is not known if it could be lowered, except for
the 3-dimensional case. Einstein 3-dimensional metrics are necessarily
of constant curvature, and thus a 3 -d Einstein metric is locally that
of a 3-sphere, flat ${I\!\!R} ^{3}$, or hyperbolic 3-space. There
 exist simple
local isometric embedings of $S^{3}\rightarrow {I\!\!R} ^{4}$ and of
 $H^{3}\rightarrow {I\!\!R} ^{5}$.
\subsection{Einstein Metrics on Three Manifolds}
To see what makes $S^{1}\times S^{2}$ very peculiar, we review here ,
again following Besse, the important mathematical facts about 3-d
Einstein manifolds.

Perhaps, the most important truth is a theorem due to R.S. Hamilton
\cite{ham82} :
{\it Let M be a connected, compact (without boundary), smooth,
 3-d manifold,
 and assume that M admits a metric g such that Ric(g) is everywhere
 positive definite. Then M also admits a metric with constant positive
 sectional curvature.}

The analogue of the Hamilton theorem for negative Ricci curvature
does not exist. The counter-example is precisely $S^{1}\times S^{2}$,
which Gao and Yau \cite{gy} have shown to possess a strictly negative Ricci
 curvature metric.
However it cannot have a negative sectional curvature metric since than
using the exponential map its universal cover would have to be
 ${I\!\!R} ^{3}$ and not ${I\!\!R}\times S^{2}$.

Since in dimension 3 an Einstein manifold has necessarily constant
sectional curvature, its universal covering is diffeomorphic either to
${I\!\!R} ^{3}$ or to $S^{3}$. Therefore $S^{1}\times S^{2}$ has
 no Einstein
 metric. For 3-manifolds admitting a metric with constant sectional
 curvature, any embedded 2-sphere in M bounds an embedded 3-ball
 $B^{3}$ in M, (M is prime). This is impossible for the membrane, and
 so is for any manifold with a non trivial connected sum decomposition
 M = N$\#$P (where N and P are not diffeomorphic to $S^{3}$).
\section{Einstein-Weyl spaces}
\subsection{3-dimensional Einstein-Weyl spaces}
In 1989 K.P. Tod \cite{t90} has compactly reviewed 3-dimensional
 Einstein-Weyl spaces. Here we follow his extremely clear exposition.
A Weyl space is a smooth (real or complex) manifold equipped with:

(1) a conformal metric

(2) a symmetric connection or torsion-free covariant derivative
(so-called Weyl connection)

which are compatible in the sense that the connection preserves the
conformal metric. This compatibility ensures two fundamental facts.
The first one is ``teleorthogonality", that is orthogonal vectors stay
orthogonal when parallel propagated in the Weyl connection. The second
is the uniqueness of the null geodesics. Once given a conformal metric
one could define null geodesics, and they preserve the null character
with respect to the Weyl connection.

These two remarkable properties have been exploited by Weyl in 1918
 to unify the
 long range fields of nature, the electromagnetic and the
 gravitational fields \cite{weyl}. In the historical perspective,
 Weyl tried to
 generalize the notion of parallel transport in general relativity to
 include the possibility that lengths, and not only directions, may
 change under parallel transport of vectors along any path. The Weyl
 theory, which made Weyl spaces interesting from the physical
standpoint, provided a geometrical interpretation for the
 electromagnetic field, but the absence of an
``absolute standard of length" has been considered a crucial
 contradiction to known experimental facts.

Coming back to mathematics and writing a chosen representative for the
conformal metric in local coordinates as $g_{ab}$, and the Weyl
covariant derivative as $D_{a}$, the compatibility condition has the
form

\begin{equation}
D_{a}g_{bc}=\omega_{a} g_{bc}
\end{equation}
for some 1-form $\omega =\omega_{a} dx^{a}$.

The conformal change of the metric
\begin{equation}
g\rightarrow \hat{g} =\Omega ^{2} g
\end{equation}
brings in the following change of the one-form
\begin{equation}
\omega\rightarrow \hat{\omega}=\omega +2\frac{d\Omega}{\Omega}
\end{equation}
where $\Omega$ is a smooth, strictly positive, function on the
 Weyl space.

The difference between the Weyl connection and the Levi-Civita connection is 
encoded in the 1-form $\omega$. Thus a Weyl space can
be defined as the pair (g,$\omega$) with $\omega$ constrained by Eq.(4).

The Weyl connection has a curvature tensor, and by contraction, a
Ricci tensor. The skew part of the Ricci tensor is a 2-form which is
automatically a multiple of $d\omega$. In order to impose the Einstein
condition on the Weyl space we constrain the symmetric part of the Ricci
W-tensor to be proportional to the conformal metric. In local
 coordinates, the Einstein condition in Weyl space reads
\begin{equation}
  W_{(ab)}=\Psi g_{ab} \;\;\;  $some$\;\; \Psi
\end{equation}
 Eq.(5) is the definition of Einstein-Weyl spaces. Moreover,
  in the 3-dimensional case one may prove the identity
  \begin{equation}
  W_{[ab]}=\;\;-3\;F_{ab}
  \end{equation}
  where $F_{ab}$ is the 2-form d$\omega$.

The Einstein condition, Eq.(5), can be rewritten in terms of
$g_{(ab)}$ and $\omega_{a}$ in the following way
\begin{equation}
R_{ab}-\frac{1}{2} \nabla _{(a}\omega _{b)} -\frac{1}{4}\omega _{a}
\omega _{b} \propto \;g_{ab}
\end{equation}
Eq.(7) shows that the Einstein-Weyl equation, Eq.(5),
is a natural conformally-invariant equation which generalises the
Einstein equation. At the same time, Eq.(7) and Eq.(6) define the 
membrane $S^{1}\times S^{2}$.
3-dimensional E-W spaces have been studied for the first time by
Cartan in 1943; he showed that 3-d E-W spaces are specified in terms
of four arbitrary functions of two variables.
\subsection{Mini-twistors and Geodesics}
Other mathematical facts which may be of physical relevance are
Hitchin description of 3-d E-W spaces in terms of mini-twistors, i.e.,
two-dimensional complex manifolds containing a family of rational
 curves \cite{h82}. Hitchin correspondence was
  extensively used by Pederson and
 Tod to construct new series of E-W spaces \cite{pt88}.
 Recently, Merkulov constructed a supersymmetric generalization
 of E-W geometry, \cite{m91}.
 The E-W structure of $S^{1}\times S^{2}$ can be obtained in an easy
 way from ${I\!\!R}^{3}$ taken as an E-W space with :
 \begin{equation}
 h=dr^{2} +r^{2}(d\theta ^{2}+\sin^{2}\;\theta d\phi ^{2})
\end{equation}
\begin{equation}
\omega =0
\end{equation}
Put $r=e^{\chi}$ and rescale with $\Omega=e^{-\chi}$:
\begin{equation}
\hat h =d\chi ^{2} +d\theta ^{2} +sin^{2}\theta d\phi ^{2}
\end{equation}
\begin{equation}
\hat \omega =-2d\chi
\end{equation}
Take $\chi$ to be periodic. Then one will find a compact E-W space with
Weyl scalar nought, $W=0$, and $F_{ij}=0$. It is $S^{1}\times S^{2}$.

Finally the most ``physical" property is the behaviour of geodesics.
The membrane could be considered to be the space between two concentric
spheres in the flat 3-d Euclidean space, with a suitable identification
of the two bounding spheres. When a geodesic hits the exterior sphere
at a certain point, it will reappear on the interior sphere at a point
on the same radial direction. It follows that the geodesic
will end up on the radial line in its plane to which it was initially
parallel.

\section{Topology Change and the $S^{1} \times S^{2}$ Membrane}

There is some progress concerning topology changes in general relativity
which is of direct interest for the hadronic membranes \cite{hgi}. The
old result of Geroch \cite{ger} that a spacetime contains at least one
closed timelike curve iff its boundary is purely spacelike or purely
timelike has been extended by Gibbons and Hawking to boundaries with
non-zero gravitational kink number. Actually, there is a lively discussion
concerning the definition of the kinking number as related to the
topological conservation laws at spacetime boundaries and to the causality
requirements \cite{low}. If the hadronic membranes are considered not
to be entirely spacelike or timelike than the topological law of
Gibbons and Hawking states that the sum of their wormhole and the number
of kinks is conserved modulo two. Furthermore, the
supersymmetric supergravity domain wall space-times considered by
Cveti\v{c} {\em et al} \cite{cv} should be taken into account as another
language for hadronic membranes and it would be interesting to see the
results. Gibbons conjectured that there may be zero- modes
trapped on the domain wall that are related to the Dirac singletons
\cite{g2}. Also the connection with the Gauss-Mainardi-Codazzi formalism
is an open issue.

\section{Conclusions}
We have considered in some detail, following leading mathematicians,
 the Einstein-Weyl character of
the $S^{1}\times S^{2}$ membrane that one has to study in the case of
Anti-de Sitter bags. This membrane happens to be a beautiful
 mathematical treasure,
and certainly is full of physical consequences, that we have yet to
explore. We recall here that de Sitter bubble-like extended models
 for the electron and
the muon have been put forth by Dirac thirty years ago \cite{d62}.
However, we do not favor too much Dirac extended models, as we consider
the electron to be similar to a disclination in a crystal.
As for hadrons endowed with $S^{1}\times S^{2}$ membranes, it will be
of interest to investigate dynamical problems, e.g., hadron-hadron
interactions at the level of their membranes.
Topological methods under current progress will probably clarify more
the mathematical structure of the hadronic membranes.

%%%%%%%%%%%%%%%%%%%%%%%%%%%%%%%%%%%%%%%%%%%%%%%%%%%%%%%%%%%%%%%%%%%%%%%%%%
\section*{}

We thank Professor Abdus Salam, the International
Atomic Energy Agency and the United Nations Educational,
Scientific and Cultural Organizations for their hospitality
at the International Centre for Theoretical Physics in Trieste, where
this work has been started. Partial support through the CONACyT Grant No.
F246-E9207 to the University of Guanajuato is gratefully acknowledged.
%%%%%%%%%%%%%%%%%%%%%%%%%%%%%%%%%%%%%%%%%%%%%%%%%%%%%%%%%%%%%%%%%%%%%%%%%%%%

\newpage
%%%%%%%%%%%%%%%%%%%%%%%%%%%%%%%%%%%%%%%%%%%%%%%%%%%%%%%%%%%%%%%%%%%%%%%%%%%%%


\begin{thebibliography}{99}
\bibitem{ro92} H. Rosu, ``Hybridizing the Skyrmion with an
 Anti- de Sitter Bag",
Nuovo Cimento B {\bf 108}, 313 (1993);

\bibitem{nnz85} S. Nadkarni, H.B. Nielsen, and I. Zahed, ``Bosonization
Relations as Bag Boundary Conditions", {\em Nucl. Phys.} B {\bf 253},
 308 (1985)

\bibitem{ff} M. Flato and C. Fronsdal, ``Quarks or Singletons",
{\em Phys. Lett} B {\bf 172}, 412 (1986)

\bibitem{t} W. Thurston, The Geometry and Topology of 3-Manifolds,
Notes based on a course given at Princeton University in 1978-1979,
(PUP, 1981)

\bibitem{be87} A.L. Besse, {\em Einstein Manifolds}, Springer Verlag, 1987

\bibitem{g} V. Guillemin ``Cosmology in (2+1)-Dimensions, Cyclic Models,
and Deformations of $M_{2,1}$", Annals of Math. Studies {\bf 121} (PUP,
Princeton, New Jersey, 1989)

\bibitem{dk81} D. DeTurck and J. Kazdan, ``Some Regularity Theorems in
Riemannian Geometry", {\em Ann. Sc. Ecole Normale Superiore},
4e. Serie 14, 249, (1981)

\bibitem{ham82} R.S. Hamilton, ``3-Manifolds with Positive Ricci
Curvature", {\em J. Diff. Geometry} {\bf 17}, 255 (1982)

\bibitem{gy} L.Z. Gao and S.T. Yau, ``The Existence of Negatively Ricci
Curved Metrics on 3-Manifolds", {\em Inv. Math.} {\bf 85}, 637 (1986)

\bibitem{weyl} H. Weyl, {\em Sitzungsber. Preuss. Akad. Wiss.}, 465 (1918)

\bibitem{t90}
 K.P. Tod, ``3-Dimensional Einstein-Weil Geometry",
 in London Math. Soc. Lect. Note Series. {\bf 150} (1990)

\bibitem{h82}
 N.J. Hitchin, ``Complex Manifolds and Einstein's
 Equations", in Lect. Notes in Math. {\bf 970} (1982)

\bibitem{pt88}
 H. Pedersen and K.P. Tod, ``3-Dimensional Einstein-Weyl
 Geometry", {\em Adv.Math} {\bf 97}, 74 (1993)

\bibitem{m91}
 S.A. Merkulov, ``Twistor Transform of Einstein-Weyl
  Superspaces", {\em Class. Quantum Grav.} {\bf 8}, 2149 (1991);
For a recent review of twistor field theory see:
S.A. Huggett, ``Recent work in twistor field theory",
{\em Class. Quantum Grav.} {\bf 9}, 127 (1992);
For the connection between minitwistor spaces and Einstein-Weyl spaces
see:
P.E. Jones and K.P. Tod, ``Minitwistor spaces and Einstein-Weyl spaces",
{\em Class. Quantum Grav.} {\bf 2}, 565 (1985)

\bibitem{d62}
 P.A.M. Dirac, ``An Extensible Model of the Electron",
 {\em Proc. R. Soc. London} A {\bf 268}, 57 (1962);
See also:
W.R. Wood, G. Papini, ``Breaking Weyl invariance in the interior of a
bubble", {\em Phys. Rev.} D {\bf 45}, 3617 (1992)

%%%%%%%%%%%%%%%%%%%%%%%%%%%%%%%%%%%%%%%%%%%%%%%%%%%%%%%%%%%%%%%%%%%%%%%%%%%%%%

\bibitem{hgi}
G.W. Gibbons and S.W. Hawking, ``Kinks and Topology Change'',
{\em Phys.Rev. Lett.} {\bf 69}, 1719 (1992); see also: A. Chamblin and
R. Penrose, {\em Twistor Newsletter} {\bf 34}, 13 (1992)

\bibitem{ger}
R.P. Geroch, ``Topology in General Relativity",{\em J. Math. Phys.} {\bf 8},
782 (1968)

\bibitem{low}
R.J. Low, ``Kinking number and the Hopf index theorem", {\em J. Math. Phys.}
{\bf 34}, 4840 (1993)

\bibitem{cv}
M. Cveti\v{c} et al., ``Cauchy Horizons, Thermodynamics, and Closed Timelike
Curves in Planar Supersymmetric Spaces",
{\em Phys. Rev. Lett.} {\bf 70}, 1191 (1993)

\bibitem{g2}
G.W. Gibbons, ``Global structure of supergravity domain wall space-times",
{\em Nucl. Phys. B} {\bf 394}, 3 (1993)

%%%%%%%%%%%%%%%%%%%%%%%%%%%%%%%%%%%%%%%%%%%%%%%%%%%%%%%%%%%%%%%%%%%%%%%%%%%%
%A. Ferrando and V. Vento,
%``Dynamical confinement in bosonized 2-d QCD",
%PR D {\bf 49}, 3044 (1994)
%
%A. Ballesteros et al.,
%``Quantum str. of the motion groups of the 2-d
%Cayley-Klein geometries", J. Phys. A {\bf 26}, 5801 (1993)
%
%E.H. Kerner, ``A novel mass-eigenvalue pb for spinors in deSitter space",
%J. Math. Phys. {\bf 25}, 150 (1984)
%
%H.V. Fagundes,
%``Smallest universe of negative curvature", PRL {\bf 70},
%1579 (1993)
%
%E. Van Beveren, T.A. Rijken, and C. Dullemond, ``Dielectric description of
%hadrons with AdS symmetry", J. Math. Phys. {\bf 27}, 1411 (1986)
%
%D.J. Navarro and J. Navarro-Salas,
%``A relativistic harmonic oscillator
%simulated by an AdS background", hep-th/9406001
%
%A. Barnes, ``Projective collineations in Einstein spaces", Class. Quantum
%Grav. {\bf 10}, 1139-45 (1993)
%
%I. Bloch and H. Crater, ``Lorentz-invariant potentials and the
%non-relativistic limit", AJP {\bf 49}, 67 (1981)
%
%J. Finger, J.E. Mandula, and J. Weyers,
%``The pion in QCD", PL B {\bf 96}, 367 (1980)
%
%S.J. Stainsby and R.T. Cahill,
%``Is space-time euclidian ``inside" hadrons ?",
%Phys. Lett. A {\bf 146}, 467 (1990);
%R.T. Cahill, ``Hadronic laws from QCD",
%Nucl. Phys. A {\bf 543}, 63c-78c (1992) [letter of Nov. 22, 1994 from Cahill]

%U. Yurtsever,
%``A remark on Kinks and Time Machines",
%Gen. Rel. Grav. {\bf 27}, 691 (1995)

%A. Chamblin, G.W. Gibbons, A.R. Steif, PR D {\bf 50}, R2353 (1994)

%M. De Francia, H. Falomir, E.M. Santangelo,
%``CC Scenario in a 3+1 dimensional Hybrid Chiral Bag", hep-ph/9507347

%S. Holst,
%``Gott time machines in AdS space",
%Gen. Rel. Grav. {\bf 28}, 387-403 (1996) [12 refs]

\end{thebibliography}
\end{document}